\documentclass[journal]{IEEEtran}
\usepackage{amsmath,amsfonts}
\usepackage{algorithmic}
\usepackage{algorithm}
\usepackage{array}
\usepackage{textcomp}
\usepackage{stfloats}
\usepackage{url}
\usepackage{verbatim}
\usepackage{graphicx}
\usepackage[compatibility=false]{caption}
\usepackage{cite}
\usepackage{subcaption}
\usepackage{xcolor}
\usepackage{hyperref}
\newcommand{\mh}[1]{\textcolor{black}{#1}}

\usepackage{nomencl} 
\makenomenclature

\begin{document}

\title{Quantifying and Improving the Accuracy of Electromagnetic Transient-Transient Stability Hybrid Simulation}

\author{
Bin Wang, \IEEEmembership{Senior Member, IEEE}, Qiang Zhang, \IEEEmembership{Senior Member, IEEE}, Xiaochuan Luo, \IEEEmembership{Fellow, IEEE}, Slava Maslennikov, \IEEEmembership{Fellow, IEEE}, Mingguo Hong, \IEEEmembership{Member, IEEE}, Xinghao Fang, \IEEEmembership{Member, IEEE}, Tongxin Zheng \IEEEmembership{Fellow, IEEE}

\thanks{The authors are with the Department of Advanced Technology Solutions, ISO New England, 1 Sullivan Rd, Holyoke, MA, 01106. Emails: \{bwang, qzhang, xluo, smaslennikov, mhong, xfang, tzheng\}@iso-ne.com.}

}



\maketitle

\footnotetext{DOI: \href{https://doi.org/10.1109/TPWRS.2026.3683988}{10.1109/TPWRS.2026.3683988}}

\begin{abstract}
The increasing penetration of inverter-based resources introduces new dynamic challenges to modern power grids, such as sub- and super-synchronous oscillations and other faster dynamics. These dynamics are typically fast in nature and are difficult to accurately model and analyze using standard transient stability (TS) methods, necessitating the need for electromagnetic transient (EMT) analysis. However, EMT simulations are notoriously slow for large-scale grids due to both equation formulations and computational limitations. To overcome this challenge, EMT-TS hybrid simulation is often used, since it offers a balanced trade-off between accuracy and speed, making it feasible to perform EMT analysis on large systems. One open question about EMT-TS hybrid simulation is the accuracy of the EMT-TS boundary or interface. This paper introduces an error index to quantify EMT-TS hybrid interface errors, identifies conditions where the hybrid simulation approach may become inaccurate, and suggests EMT region expansions to improve the simulation accuracy. Additionally, a three-sequence hybrid interface model is proposed to mitigate inaccuracies caused by unbalanced conditions.
\end{abstract}


\printnomenclature
\nomenclature{$V_{\text{m,h\_{emt,pu}}}(t)$}{Per-unit voltage magnitude trajectory of the boundary bus measured from the EMT side of the hybrid simulation}
\nomenclature{$V_{\text{m,full\_{emt,pu}}}(t)$}{Per-unit voltage magnitude trajectory of the boundary bus measured from the full EMT simulation}
\nomenclature{$e_{\text{true}}$}{True error of the hybrid simulation defined in (\ref{eq:etrue})}
\nomenclature{$e_{\text{idx}}$}{The proposed error index reflecting the true error of the hybrid simulation defined in (\ref{eq:eidx})}
\nomenclature{$\Delta V_{\text{diff}} (t)$}{Difference in three-phase instantaneous voltages at the boundary bus between the EMT and TS sides of the hybrid simulation}
\nomenclature{$V_{\text{B}}$}{Base line-to-line voltage at the boundary bus}
\nomenclature{$V_{i \text{,h\_emt}}(t)$}{Three-phase instantaneous voltages in physical unit directly measured from the EMT side of the hybrid simulation}
\nomenclature{$V_{i \text{,h\_ts}}(t)$}{Three-phase instantaneous voltages in physical unit reconstructed from TS-side per-unit voltage phasor using using (\ref{eq:vabc_ts})}
\nomenclature{$f_0$}{Nominal fundamental frequency}
\nomenclature{$V_{\text{m,h\_{ts,pu}}}(t)$}{Per-unit voltage magnitude trajectory of the boundary bus measured from the TS side of the hybrid simulation}
\nomenclature{$\theta_{\text{h\_{ts}}} (t)$}{Voltage angle trajectory (in rad) of the boundary bus measured from the TS side of the hybrid simulation}
\nomenclature{$e^\prime_{\text{idx}}$}{Modified error index defined in (\ref{eq:eidx_int_mod})}

\section{Introduction}
\label{sec:intro}

The rapid growth of \mh{power electronic} inverter-based resources (IBR) \cite{2012chen:} has introduced unprecedented reliability risks to \mh{the} US power grids, \mh{as findings from}
the 2016 Blue Cut Fire 
\cite{nerc2017} and 2022 Odessa disturbances 
\cite{nerc2022_odessa}
\mh{investigation reports strongly suggest. There are emerging issues with IBR dynamics that are inherently fast,} and it is well \mh{understood} that \mh{the traditional} transient stability (TS) simulation \mh{tools cannot fully capture the fast} IBR dynamics \mh{including both the sub- and super-synchronous oscillation (SSOs)} at \mh{frequencies $7$ Hz and above} \cite{2022survey:sso}. To address this \mh{critical gap}, the North American Electric Reliability Corporation (NERC) recommends detailed electromagnetic transient (EMT) simulation \mh{studies} \cite{nerc2022_emt,2012fan:SFA}. However, widespread adoption of EMT simulation in large-scale power system planning and operation \mh{studies has not yet been realized; in addition to the limited workforce and trainings on EMT in the industry, it is also} hindered by significant computational challenges \mh{such as} i) the large-scale integration of \mh{high-order IBR models}, ii) \mh{requirement} for small \mh{numerical integration} time steps (10-50 $\mu$s or smaller) and iii) high dimensionality of the \mh{power system} network. Novel modeling and simulation techniques, such as the heterogeneous multiscale method \cite{huang2025} and dynamic phasor method \cite{fan2012}, are emerging as promising approaches. However, their fast simulation capabilities are not yet widely available in commercial tools nor broadly adopted in practice, as these methods are still progressing toward full maturity and commercialization.

Given the growing urgency of \mh{industry's need for} EMT studies, EMT-TS hybrid simulation has emerged as a promising strategy to \mh{facilitate large-scale applications as} highlighted by \cite{2023luo:ISONE}, \cite{2024zhang:ISONE} and \cite{2024dong:opalrt}. Specifically, EMT-TS hybrid simulation has been incorporated into routine Operations Studies at ISO-NE since 2023. By modeling most \mh{part} of the system in \mh{the} TS simulation while \mh{keeping selected components --- such as IBRs and transmission lines that drive dynamic instabilities --- in the more accurate EMT simulation,} this approach offers a practical \mh{strategy to} balance between accuracy and speed \cite{2024dong:D,2024dong:gfm}. \mh{Among the previous development efforts, reference} \cite{2024xiong:hybrid} introduced an EMT-TS hybrid simulation framework that integrates the open-source EMT solver ParaEMT \cite{paraemt2024} with the TS solver GridPACK via HELICS. Commercially available EMT-TS hybrid interfaces include HYPERSIM-ePHASORSIM, PSCAD-TSAT, RTDS-TSAT and PSCAD-PSSE. 

\mh{During EMT-TS hybrid simulations, a number of network buses/branches are designated as the boundary so that sub-network models can be created for EMT and TS. Incremental exchange of boundary state signals takes place between EMT and TS throughout the simulation time steps. Most EMT-TS interfaces} perform well when \mh{the} signals exchanged consist solely of fundamental-frequency positive-sequence components. This is a reasonable assumption from \mh{the} design perspective because (i) \mh{the TS model is built to represent the fundamental-frequency dynamics of the positive sequence network and (ii) the network boundary buses of the hybrid simulation interface should be chosen to support the modeling assumption of (i)}. \mh{In practice, however, the hybrid simulation interface buses may be chosen based on many} engineering considerations \mh{such as} estimated size of the EMT region \mh{and the} critical fault locations, etc., without \mh{advance knowledge if the signals exchanged across the interfaces will contain more than the fundamental-frequency and positive sequence components.} For example, reference \cite{Khamees2025} reported that EMT-TS hybrid simulations can become inaccurate when \mh{a} significant amount of harmonics or unbalanced conditions appear near an interface bus. To assess \mh{the} inaccuracy, reference \cite{Khamees2025} proposed two indices, Harmonic Magnitude and Phase Imbalance. \mh{These indices are proposed to measure the extent of violations concerning the assumptions (i) and (ii).} However, their calculations rely on the availability of the \mh{full results of} EMT simulation --- data that \mh{can become} unavailable when the system scale is large. \mh{Meanwhile, }the interface errors \mh{are also related to the fault locations and the study scenario configurations}. \mh{Although the proposed indices are theoretically sound, they are only useful for the study of small systems and become impractical for large-scale applications.}

\mh{So far, there have been few reported studies on EMT-TS hybrid simulation and especially on the criteria for choosing the interface or boundary buses \cite{Khamees2025,2024xiong:hybrid,2024mahsa:hybrid,2024epri:hybrid}. In practice, engineers make an empirical judgment in choosing the initial boundary and make adjustments if necessary.}

The primary motivation of this paper is to continue the research and develop a systematic yet still practical approach for determining the hybrid simulation boundary, minimizing computational complexity while preserving accuracy. A prerequisite for achieving this goal is the ability to assess the accuracy of an EMT-TS hybrid simulation for any given boundary. This paper introduces a first-of-its-kind error index for hybrid simulation eliminates the need for full EMT simulations, offering a promising tool for practical boundary determination. To this end, this paper 

\begin{itemize}
\setlength\itemsep{0em}
\item Proposes an error index to calculate the hybrid interface's accuracy using only signals obtained from the hybrid simulation, eliminating the need for full EMT simulation.
\item Benchmarks the proposed error index against the true error of hybrid simulation and demonstrates its effectiveness.
\item Identifies most important factors affecting hybrid simulation accuracies and proposes a systematic approach in adjusting boundaries to improve the hybrid simulation accuracy.
\item Introduces a three-sequence hybrid interface model to mitigate inaccuracies caused by unbalanced faults when boundary adjustments are impractical.
\end{itemize}

Note that this work assumes a bus-type hybrid boundary, where boundary buses are modeled in both EMT and TS domains, similar to the PSCAD-PSSE ETran Plus interface by \cite{2016electranix:etranplus}. However, the proposed approach may be generalized to include branch-based boundaries such as the PSCAD-TSAT interface, where one end of the branch is retained in PSCAD and the other end in TSAT, while the branch itself can be retained in either PSCAD or TSAT per user's choice \cite{2020powertech:tpi}.

The remainder of the paper is structured as follows: Section \ref{sec:true_er} defines the true error; Section \ref{sec:eridx} introduces the proposed error index; Section \ref{sec:num_faults} presents numerical tests on a small 4-bus system using both balanced and unbalanced faults; Section \ref{sec:num_FOs} extends the analysis to account for other factors affecting hybrid simulation accuracies, including forced oscillations; Section \ref{sec:3seq} introduces a three-sequence interface model to enhance the accuracy of hybrid simulations for unbalanced fault studies, and Section \ref{sec:con} concludes the paper.

\section{True Error of Hybrid Simulation}
\label{sec:true_er}

To evaluate the accuracy of hybrid simulation, a reference is required to represent the ground truth of the hybrid simulation error. In this work, the full EMT model and simulation serve as the reference. Note that this is only used to test and validate the proposed error index at the research stage, and is not needed when using the error index.

\subsection{Modeling Error Versus Interface Error}
Theoretically, hybrid simulation errors can be put in two categories: 

\begin{itemize}
\setlength\itemsep{0em}
\item \textbf{Modeling error (Type-1):} This type of error arises from simplifying a portion of the system in the TS domain, where the discrepancies are caused by modeling differences between EMT and TS.
\item \textbf{Interface error (Type-2):} This type of error results from information loss at the hybrid interface, specifically how accurately and timely signals are exchanged between the EMT and TS solvers.
\end{itemize}

This study focuses exclusively on Type-2 errors, assuming that there are no Type-1 errors. Because Type-1 errors are largely independent of the hybrid interface itself and should be considered separately when deciding the EMT-TS boundary (i.e., the size of the EMT region). In particular, if the EMT-relevant dynamics exist on the TS side near the interface and are of interest to the study, the corresponding components should be modeled in EMT and moved to the EMT side since it leads to Type-1 modeling errors. Type-1 errors can also be minimized by improving the modeling of dynamic components in the TS domain to better account for faster dynamics. The proposed error index is not intended to quantify such modeling errors. Instead, it tries to represent the errors introduced by the hybrid interface itself. The test system is also set up to ensure that there are no significant Type-1 errors introduced to the studies.

\subsection{Definition of True Error}
In general, the error of a hybrid simulation can be quantified by the difference between the hybrid simulation and its corresponding full EMT simulation, and it can be defined on any common electrical quantity. For the bus-type hybrid boundary and the specific commercial tools used in this study---PSSE, PSCAD, and the ETran Plus interface---the error of the hybrid simulation is defined as follows.

\begin{figure}[tb!]
    \centering

    \begin{subfigure}[b]{1.0\linewidth}
        \centering
        \includegraphics[width=0.5\linewidth]{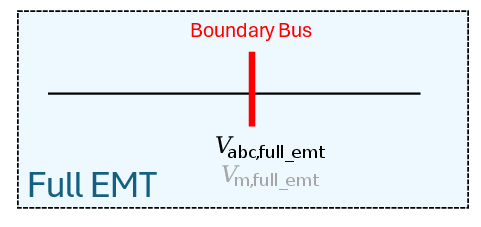}
        \caption{Full EMT model}
        \label{fig:BoundaryBus_emt}
    \end{subfigure}

    \vspace{15pt} 

    \begin{subfigure}[b]{1.0\linewidth}
        \centering
        \includegraphics[width=0.5\linewidth]{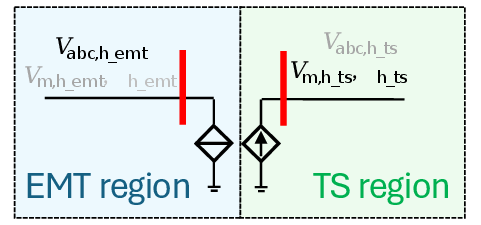}
        \caption{EMT-TS Hybrid model}
        \label{fig:BoundaryBus_hybrid}
    \end{subfigure}

    \caption{Boundary bus and measured quantities}
    \label{fig:BoundaryBus}
\end{figure}

Consider a boundary bus and the electrical quantities measured at this bus in both the full EMT and hybrid simulations, as shown in Figure \ref{fig:BoundaryBus}. In this figure, quantities with black subscripts represent differential or algebraic electric variables, while those with gray subscripts correspond to sensor-based measurements or reconstructed signals. The \textbf{true error} of the hybrid simulation, denoted as $e_{\text{true}}$, is defined as the definite integral of the voltage phasor magnitude difference over a selected time window:

\begin{equation}
    e_{\text{true}} = \int_{t_{\text{start}}}^{t_{\text{end}}} {\|V_{\text{m,h\_{emt,pu}}}(t) - V_{\text{m,full\_emt,pu}}(t)  \| dt}. \label{eq:etrue}
\end{equation}

\noindent where $V_{\text{m,h\_{emt,pu}}}(t)$ is the per-unit voltage magnitude trajectory of the boundary bus measured from the EMT side of the hybrid simulation; $V_{\text{m,full\_{emt,pu}}}(t)$ is the same quantity measured from the full EMT simulation; $t_{\text{start}}$ and $t_{\text{end}}$ respectively represent the start and end times of the window used for computing the error index; and $\|\cdot\|$ denotes the $L^2$-norm and the absolute value in the case of a scalar function. Note that in practice, the index is computed using simulated discrete samples of these trajectories, in which case the $l^2$-norm is employed as a numerical approximation of the $L^2$-norm. 

The rationale for defining error this way is outlined below.

\begin{itemize}
\setlength\itemsep{0em}
\item \textbf{Choosing boundary bus quantities instead of internal EMT-region quantities:} While quantities inside the EMT region are available in both full EMT and hybrid simulations, their selection may depend on specific cases, limiting generalizability. In contract, using boundary bus quantities ensures a more consistent and universally applicable approach.
\item \textbf{Choosing voltage instead of current or power:} The magnitude of current and power vary significantly depending on system conditions, making them less reliable for general error quantification. In contrast, voltage typically stabilizes around  $\mathbf{1.0}$ \textbf{pu} in steady-state operation, making voltage-based errors more broadly applicable.
\item \textbf{Choosing phasor magnitude instead of waveform:} Although three-phase waveforms theoretically provide the most accurate representation---free from distortions introduced by phasor calculations and capable of capturing harmonics and DC components---practical challenges arise. Full EMT and hybrid simulation operate as separate AC systems, and due to slight differences in initialization, their pre-disturbance steady states may not be perfectly phase-aligned, even if their frequencies are synchronized. To mitigate this issue, this study defines error based on voltage phasor magnitude. However, error definitions based on waveforms remain feasible if disturbance timing is carefully controlled to align waveform phases between the two simulations.
\item \textbf{Choosing $V_{\text{m,h\_emt}}$ instead of $V_{\text{m,h\_ts}}$:} In an ideal hybrid simulation, the voltage phasor magnitude measured in the EMT region should be identical to, or at least closely match, that is measured in the TS region. However, with the PSCAD-PSSE ETran Plus interface, it's been observed that discrepancies arise when the positive-sequence phasor magnitude of the boundary voltage drops below $\mathbf{0.2-0.4}$ \textbf{pu}. Despite this, voltages measured in the EMT region (along with other EMT quantities) always remain more accurate than their counterpart in the TS region, making them the preferred reference for defining the true error of hybrid simulation in this work. Further discussion on this issue is provided in Section \ref{sec:example}.
\item \textbf{Choosing time window:} The choice of the time window should not affect the results in the absence of Type-1 error. However, when a small Type-1 error is present, using a longer time window in the calculation of the error index (as defined in Equation \eqref{eq:etrue}) may amplify its influence. To minimize the impact of Type-1 error and highlight the Type-2 error, the time window should be restricted to only encompass the relevant dynamic phenomena. In this study, a 2-second post-disturbance time window is used consistently across all cases.
\end{itemize}

\section{Proposed Error Index}
\label{sec:eridx}

The proposed error index is defined below, utilizing only information available within a hybrid simulation:

\begin{subequations}
	\begin{eqnarray}
		e_{\text{idx}} = \int_{t_{\text{start}}}^{t_{\text{end}}} \Delta V_{\text{diff}} (t) dt \;\;\;\;\;\;\;\;\;\;\;\;\;\;\;\;\;\;\;\;\;\;\;\;\;\;\;\;\;\;\;\;\label{eq:eidx_int}\\
		\Delta V_{\text{diff}} (t) = {\frac{1}{\sqrt{6} V_{\text{B}}}  \sqrt{\sum_{i}\|V_{i \text{,h\_emt}} (t) - V_{i \text{,h\_ts}} (t)\|^2}}.  \label{eq:eidx_integrand}
	\end{eqnarray}\label{eq:eidx}%
\end{subequations}

\noindent where $i\in \{a,b,c\}$, $V_{i \text{,h\_emt}}(t)$ represents the three-phase waveforms in physical unit directly measured from the EMT side of the hybrid simulation, $V_{\text{B}}$ is the base line-to-line voltage at the boundary bus, and $V_{i \text{,h\_ts}}(t)$ represents the three-phase waveforms in physical unit reconstructed from TS-side per-unit voltage phasor using (\ref{eq:vabc_ts}). The purpose of having the denominator of $\sqrt{6}V_{\text{B}}=\frac{V_{\text{B}}}{\sqrt{3}}\cdot\sqrt{2}\cdot3$ in (\ref{eq:eidx_integrand}) is to ensure that the proposed error index in (\ref{eq:eidx}) has the same unit as the true error in (\ref{eq:etrue}) and their values have comparable magnitudes, where $\frac{V_{\text{B}}}{\sqrt{3}}$ is for converting voltage from physical unit to per unit, $\sqrt{2}$ for scaling maximum instantaneous value to root-mean-square value, and $3$ is adopted here to compute the average error among three phases. Note that the proposed error index has the unit pu$\cdot$s. If desired, it can be normalized by the integration interval to represent an average voltage error. In this paper, such normalization is not applied, and the proposed index is used to quantify the total accumulated error.

\begin{subequations}
	\begin{eqnarray}
		V_{a \text{,h\_ts}} = \sqrt{\frac{2}{3}} V_{\text{B}} V_{\text{m,h\_{ts,pu}}}(t) \sin \left( 2\pi f_{0} t + \theta_{\text{h\_{ts}}} (t) \right) \;\;\;\;\;\;\;\;\;\;\;\; \label{eq:va_ts}\\
		V_{b \text{,h\_ts}} = \sqrt{\frac{2}{3}} V_{\text{B}} V_{\text{m,h\_{ts,pu}}}(t) \sin \left(2\pi f_{0} t + \theta_{\text{h\_{ts}}} (t) - \frac{2\pi}{3}\right)  \label{eq:vb_ts}\\
            V_{c \text{,h\_ts}} = \sqrt{\frac{2}{3}} V_{\text{B}} V_{\text{m,h\_{ts,pu}}}(t) \sin \left(2\pi f_{0} t + \theta_{\text{h\_{ts}}} (t) + \frac{2\pi}{3}\right)  \label{eq:vc_ts}
	\end{eqnarray}\label{eq:vabc_ts}%
\end{subequations}

\noindent where the nominal fundamental frequency $f_0$ is used to reconstruct three-phase waveforms from phasor voltages. This choice is consistent with phasor-domain (TS) modeling where network equations and phasor quantities are defined with respect to a fixed fundamental frequency. Although system dynamics may involve frequency deviations, such effects are represented in the TS simulation through the time-varying voltage phase angles instead of a time-varying frequency.

Key considerations for the proposed error index are:
\begin{itemize}
\setlength\itemsep{0em}
\item \textbf{Clear physical meaning:} If $e_{\text{idx}} = 0$, the three-phase waveforms reconstructed from the positive-sequence phasor perfectly align with the EMT-side three-phase waveforms. In this scenario, the electric quantities at the boundary are purely positive-sequence, meaning no error is introduced by the hybrid interface. If $e_{\text{idx}}$ is significantly greater than zero, it indicates the presence of substantial non-positive-sequence component that are not transmitted across the interface between EMT and TS solvers, leading to errors in the hybrid simulation.
\item \textbf{Independent of full EMT simulation:} The proposed error index is computed solely from data available in a hybrid simulation, eliminating the need for a full EMT reference. Specifically, it requires only the EMT-side three-phase voltage waveforms and the TS-side voltage phasor, both of which are locally available at the boundary bus. This ensures straightforward implementation in commercial tools such as PSCAD and ETran Plus interface model.
\item \textbf{Integration window selection:} Theoretically, the proposed error index is independent of the choices of integration intervals. In practice, however, numerical artifacts from the differences in the simulation tools and EMT-TS interface can introduce small steady-state mismatches, i.e., Type-1 error, at the boundary buses even under undisturbed conditions (see results in section \ref{sec:num_faults}-B). To remove the impact of this Type-1 error on the proposed error index, the integration window should be limited to the time periods when the hybrid-simulation-induced discrepancies are most pronounced. Accordingly, the interval is chosen to start at fault inception and extend for several seconds afterward, covering the fault-on period, activation of protection, and post-fault transients such as IBR ride-through recovery periods.
\end{itemize}


The only problem arises in situations where the interface modeling issue discussed in Section \ref{sec:true_er} leads to $V_{\text{m,h\_emt}}(t) \neq V_{\text{m,h\_ts}}(t)$ during periods of low boundary voltage. In such cases, a significant $\Delta V_{\text{diff}}(t)$ may result from an inaccurate $V_{\text{m,h\_ts}}(t)$. This can incorrectly suggest a large error in the hybrid simulation, specifically in the EMT region. An example will be illustrated in Fig. \ref{fig:BusV_ABCG} in Section IV-B. To address this, we propose a slight modification to Eq. (\ref{eq:eidx_int}) as a mitigation to detect and suppress these false large errors. Typical small and large error values are presented for a small 4-bus test case in Section \ref{sec:num_faults}-B.

\begin{equation}
    e^\prime_{\text{idx}} = \int_{t_{\text{start}}}^{t_{\text{end}}}  H\left(V_{\text{m,h\_{ts}}}(t),a\right)   \Delta V_{\text{diff}} (t) dt. \label{eq:eidx_int_mod}
\end{equation}

\begin{equation}
    H(x,a) = 
    \begin{cases}
        1 & \text{if $x>a$} \\
        0 & \text{otherwise} 
    \end{cases}
    \label{eq:hstep}
\end{equation}

\noindent where the Heaviside step function $H(x,a)$ vanishes when the input $x$, i.e., $V_{\text{m,h\_ts}}(t)$ in (\ref{eq:eidx_int_mod}), is below the given threshold $a$. For all case studies presented in this paper, we employ PSCAD-PSSE ETran Plus interface with $a=0.2$ pu. This specific number is chosen based on statistics collected in the case studies and can be adjusted by users.

\noindent \textbf{Remark:} Sections IV and V evaluate the proposed error index through extensive fault and forced oscillation case studies, respectively. The fault studies in Section IV demonstrate that the proposed error index effectively captures hybrid simulation errors caused by the hybrid interface. In particular, large error index values, most notably under unbalanced fault conditions, indicate inaccuracies of the hybrid simulation and can be used to infer that expansion of the EMT area may be necessary. When such expansion is infeasible or undesirable, Section VI introduces a three-sequence interface model to improve hybrid simulation accuracy for unbalanced faults.

In contrast, the forced oscillation studies in Section V will demonstrate that the proposed error index can quantify hybrid simulation errors associated with forced oscillations. Such errors are caused by the interface models and . may not be effectively mitigated by expanding the EMT areaStill 


\section{Case Studies on Faults}
\label{sec:num_faults}

In the current and the following sections, extensive tests are conducted involving faults and forced oscillations to identify scenarios where hybrid simulations may become inaccurate due to the hybrid interface. Two primary situations are observed: (i) unbalanced operating conditions and (ii) oscillatory conditions. For forced oscillations that span the hybrid boundary, hybrid simulation is not recommended unless the interface modeling can be improved (also see the discussion at the end of Section~\ref{sec:num_FOs}). For unbalanced conditions, Section~\ref{sec:3seq} proposes a three-sequence interface model to accurately capture boundary behaviors.

This section demonstrates the proposed error index by comparing it with the true error of the hybrid PSCAD-PSSE simulation on a four-bus test system. Various conditions will be evaluated to (i) identify when the hybrid simulation becomes inaccurate and (ii) assess how effectively the proposed error index reflects such inaccuracies without relying on a full EMT simulation. 

\subsection{Four-Bus Test System}
\begin{figure}[tb!]
    \centering
    \includegraphics[width=\columnwidth]{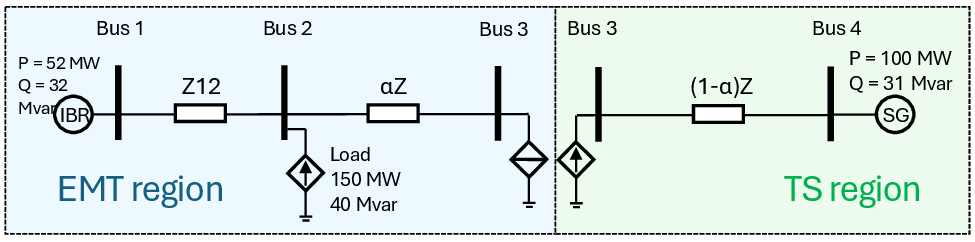} 
    \caption{Four-bus test system}
    \label{fig:FourBusDiagram}
\end{figure}

The four-bus system, depicted in Figure \ref{fig:FourBusDiagram}, has been specifically designed for this evaluation. It is simple enough to analyze, while still maintains the complexities needed to effectively test the capabilities of the EMT-TS hybrid simulation. In all test cases, Bus 3 serves as the boundary bus.

The power flow conditions are summarized in Table \ref{tab:pfss}, while the static data and dynamic data are presented in Table \ref{tab:4busdata}. Note that the choice of dynamic models does not affect the proposed error index or the conclusions of the paper that focus on the potential errors caused by the EMT-TS interface. This four-bus system features a load at Bus 2 supplied by two generators: an IBR at Bus 1 and a synchronous machine (SG) at Bus 4. Notably, the IBR is modeled using EPRI's PV-MOD-EMT-IBR v1.0 beta model with default parameters \cite{2023epri:emtibr}. The only modification made is relocating the scaling from the high side of the transformer between Bus 1 and Bus 2 to the low side. This modification ensures that the full EMT and hybrid steady-state power injections from the IBR at Bus 1 align with the specified power flow conditions.

\begin{table}[tb!]
	\centering
        \small
	\caption{Power Flow}
	\label{tab:pfss}
	\begin{tabular}{c|c|c|c|c|c}
		\hline
		Bus    & Bus  & $P_{\text{inj}}$ & $Q_{\text{inj}}$ & $V_{\text{m}}$  & $\theta$  \\ 
          \#  &  Type &  in MW & in Mvar & in pu & deg  \\ \hline
		1  & Slack     & -  & - & 1.02 & 0.0 \\ 
		2  & PQ     & -150  & -40 & - & - \\ 
            3  & PQ     & 0  & 0 & - & - \\ 
            4  & PV     & 100  & - & 1.04450 & - \\ 
        \hline
	\end{tabular}
\end{table}

\begin{table}[tb!]
    \centering
    \small
    \caption{Static and Dynamic Data}
    \label{tab:4busdata}
    \begin{tabular}{p{1.7cm}|p{5.3cm}}
    \hline
    Static Data    &  $\text{MVA}_{\text{IBR}}=100$, $\text{MVA}_{\text{SG}}=555$, Base kV $= 34.5$ except for $0.6$ for Bus 1, $Z_{12}=0.006+j0.06$ pu, Turn ratio $k_{12}=1.025$, $Z=0.002+j0.02$ pu, X/R ratio$=10$ for lines 2-3 and 3-4, $\text{SCR}_{\text{Bus2}}=5$, $\alpha=0.1$  \\ \hline
    Dynamic Data of SG  &  GENROU: $T^{\prime}_{\text{do}}=6.0$, $T^{\prime\prime}_{\text{do}}=0.5$, $T^{\prime}_{\text{qo}}=1.0$, $T^{\prime\prime}_{\text{qo}}=0.05$, $H=2.41$, $D=0.0$, $X_{\text{d}}=1.4$, $X_{\text{q}}=1.35$, $X^{\prime}_{\text{d}}=0.3$, $X^{\prime}_{\text{q}}=0.6$, $X^{\prime\prime}_{\text{d}}=X^{\prime\prime}_{\text{q}}=0.2$, $X_{\text{l}}=0.1$, $S(1.0)=0.03$, $S(1.2)=0.4$; SEXS: $T_{A}/T_{B}=0.1$, $T_{B}=10.0$, $K=100$, $T_{\text{E}}=0.1$, $E_{\text{MIN}}=0.0$, $E_{\text{MAX}}=3.0$; TGOV1: $R=0.08$, $T_{1}=2.0$, $V_{\text{MAX}}=1.0$, $V_{\text{MIN}}=0.0$, $T_{2}=3.0$, $T_{3}=15.0$, $D_{\text{T}}=0.4$  \\ \hline
    Dynamic Data of IBR  &  Default parameters in \cite{2023epri:emtibr} are used. \\ \hline
    Load & Static constant-impedance model \\ \hline

    \hline
    \end{tabular}
\end{table}

ETran is utilized to convert the system model represented by PSSE RAW and DYR files respectively into a full EMT model in PSCAD and a hybrid model in both PSCAD and PSSE. For all simulations presented, PSCAD uses a time step of 20 $\mu$s, while PSSE uses a time step of 8.33 ms (half-cycle).

\subsection{Illustration of Proposed Error Index}
\label{sec:example}

\begin{figure}[tb!]
	\centerline{\includegraphics[width=1\columnwidth]{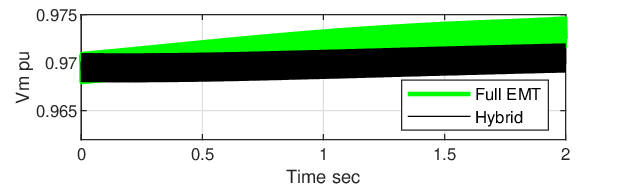}}
	\caption{No-disturbance response of bus 3 voltage}\label{fig:NoDist}
\end{figure}

\textbf{Small-error case:} The first example involves a no-disturbance simulation that verifies the steady state of the hybrid simulation against the full EMT simulation, illustrating the small magnitude of both the true error and the proposed error index in a scenario expected to yield high accuracy. In both the hybrid and full EMT simulations, after enabling all loads, generators and their controls, the system reaches a steady-state condition. A $2$-second no-disturbance response is extracted and presented in Figure \ref{fig:NoDist}. The maximum error in the positive-sequence voltage magnitude at the boundary bus is measured to be $0.0078$ pu. Over this $2$-second interval, the true error and the proposed error index are calculated as $e_{\text{true}}=0.0040$ pu$\cdot$s and $e_{\text{idx}}=0.0094$ pu$\cdot$s, respectively. These values indicate the minimal error associated with the hybrid simulation.

\begin{figure}[tb!]
    \centering
    
    \begin{subfigure}[b]{1.0\linewidth}
        \centering
        \includegraphics[width=0.9\linewidth]{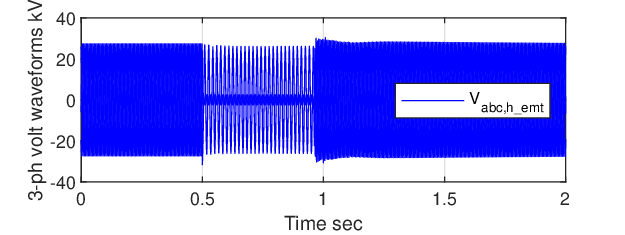}
        \caption{Three-phase waveforms measured from EMT side}
        \label{fig:Vabc_pscad}
    \end{subfigure}

    \vspace{15pt} 
    
    \begin{subfigure}[b]{1.0\linewidth}
        \centering
        \includegraphics[width=0.9\linewidth]{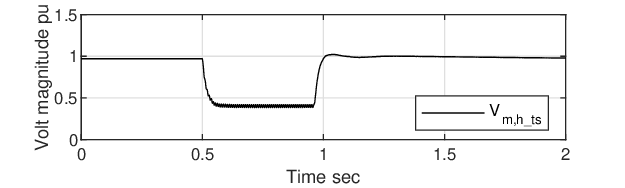}
        \caption{Positive-sequence magnitude from TS side}
        \label{fig:Vm_psse}
    \end{subfigure}

    \vspace{15pt} 

    \begin{subfigure}[b]{1.0\linewidth}
        \centering
        \includegraphics[width=0.9\linewidth]{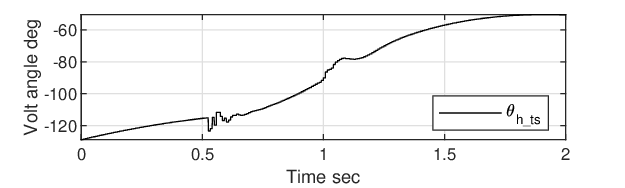}
        \caption{Positive-sequence angle from TS side}
        \label{fig:theta_psse}
    \end{subfigure}
    
    \caption{Boundary bus voltage in example 2}
    \label{fig:BusV_BCG}
\end{figure}

\begin{figure}[tb!]
	\centerline{\includegraphics[width=0.9\columnwidth]{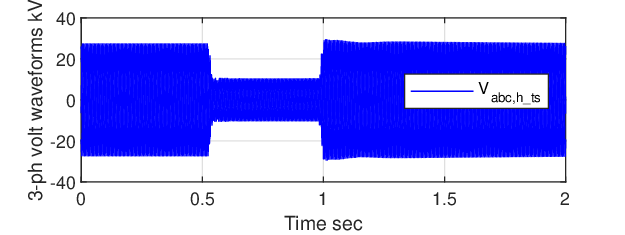}}
	\caption{Three-phase waveforms of boundary bus re-constructed from positive-sequence voltage phasor}\label{fig:Vabc_psse}
\end{figure}

\begin{figure}[tb!]
	\centerline{\includegraphics[width=0.9\columnwidth]{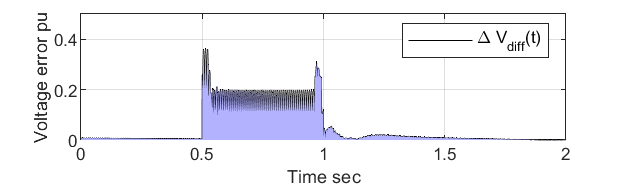}}
	\caption{Voltage difference between EMT side and TS side of the hybrid boundary}\label{fig:dVt}
\end{figure}

\begin{figure}[tb!]
	\centerline{\includegraphics[width=0.9\columnwidth]{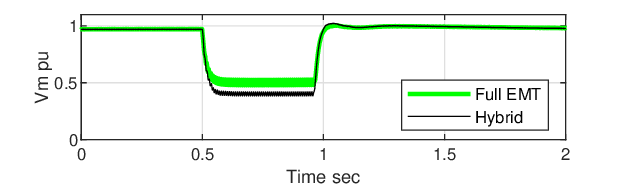}}
	\caption{Positive-sequence voltage magnitude of boundary bus in example 2}\label{fig:ex2_Vm}
\end{figure}

\begin{figure}[tb!]
	\centerline{\includegraphics[width=0.9\columnwidth]{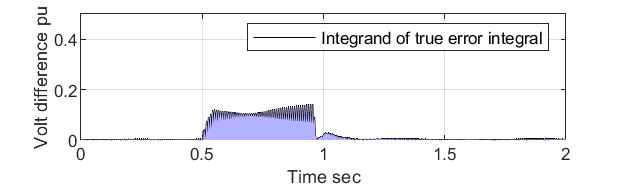}}
	\caption{True error of hybrid simulation in example 2}\label{fig:trueer}
\end{figure}

\textbf{Large-error case:} The second example demonstrates the detailed process of calculating the proposed error index using an unbalanced fault scenario. A phase-BC-to-ground fault is introduced at Bus 2 at $t=0.5$ s and cleared at $t=0.96$ s. Three-phase voltage waveforms, along with the positive-sequence voltage phasor magnitude and phase, are measured in the hybrid simulation and are shown in Figure \ref{fig:BusV_BCG}. Figure \ref{fig:Vabc_pscad} illustrates that, upon the fault's occurrence, the instantaneous voltages of the two faulted phases drop suddenly to below $0.1$ pu, while the healthy phase remains above $0.95$ pu. Figure \ref{fig:Vm_psse} indicates that positive-sequence voltage phasor magnitude only decreases to around $0.4$ pu. Using the positive-sequence phasor data in Figures \ref{fig:Vm_psse} and \ref{fig:theta_psse} to reconstruct the three-phase voltage waveforms results in Figure \ref{fig:Vabc_psse}. Calculating the difference between the waveforms in Figures \ref{fig:Vabc_pscad} and \ref{fig:Vabc_psse} according to Eq. (\ref{eq:eidx_integrand}) yields the black curve in Figure \ref{fig:dVt}. By the definition in Eq. (\ref{eq:eidx_int}), the proposed error index quantifies the area between this black curve and the horizontal axis. In this example, the calculated error index is $e_{\text{idx}} = 0.0984$ pu$\cdot$s. In comparison, the positive-sequence voltage magnitude calculated from the EMT side at the boundary bus is depicted in Figure \ref{fig:ex2_Vm}, and the true error of the hybrid simulation is found to be $e_{\text{true}} = 0.0546$ pu$\cdot$s, as illustrated in Figure \ref{fig:trueer}. These values reflect the inaccuracies present in the hybrid simulation.

\begin{figure}[tb!]
    \centering
    \begin{subfigure}{0.9\columnwidth}
        \centering
        \includegraphics[width=\linewidth]{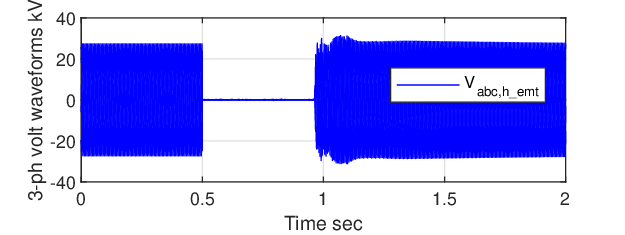}
        \caption{Three-phase waveforms measured from EMT side}
        \label{fig:ex3_Vabc_pscad}
    \end{subfigure}
    
    \vspace{15pt} 
    
    \begin{subfigure}{0.9\columnwidth}
        \centering
        \includegraphics[width=\linewidth]{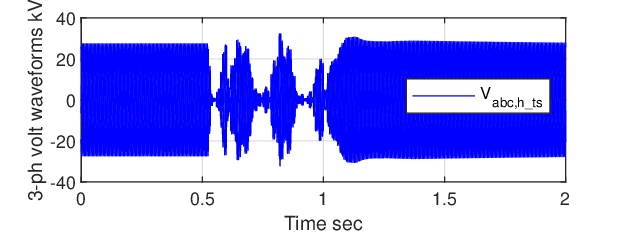}
        \caption{Three-phase waveforms re-constructed from TS side}
        \label{fig:ex3_Vabc_psse}
    \end{subfigure}
    
    \caption{Boundary bus voltage in example 3}
    \label{fig:BusV_ABCG}
\end{figure}

\begin{figure}[tb!]
    \centering
    \begin{subfigure}{0.9\columnwidth}
        \centering
        \includegraphics[width=\linewidth]{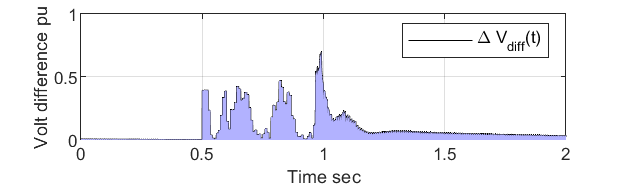}
        \caption{Voltage difference per eq. (\ref{eq:eidx_integrand})}
        \label{fig:ex3_dvt}
    \end{subfigure}

    \vspace{15pt} 
    
    \begin{subfigure}{0.9\columnwidth}
        \centering
        \includegraphics[width=\linewidth]{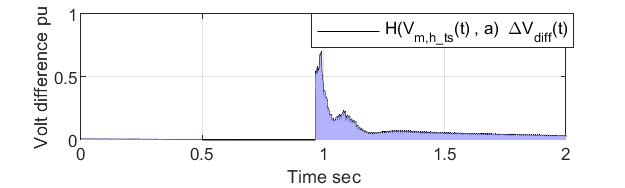}
        \caption{Voltage difference per eq. (\ref{eq:eidx_int_mod})}
        \label{fig:ex3_dvt_mod}
    \end{subfigure}
    
    \caption{Boundary bus voltage in example 3}
    \label{fig:erridx_lowV}
\end{figure}

\begin{figure}[tb!]
	\centerline{\includegraphics[width=0.9\columnwidth]{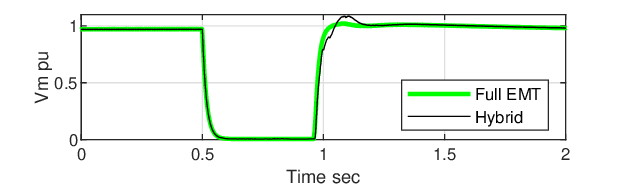}}
	\caption{EMT-side positive-sequence voltage magnitude of boundary bus in example 3}\label{fig:ex3_Vm}
\end{figure}

\textbf{Low boundary voltage case:} The third example illustrates an issue with the specific EMT-TS interface utilized in this study, namely the PSCAD-PSSE ETran Plus interface, and demonstrates how the modified error index can help mitigate this issue. A three-phase fault is introduced at Bus 3, the boundary bus, at $t=0.5$ s and cleared at $t=0.96$ s. The three-phase voltage waveforms from the EMT side of the boundary and those reconstructed from the TS side voltage phasor are shown in Figures \ref{fig:ex3_Vabc_pscad} and \ref{fig:ex3_Vabc_psse}, respectively. Voltage waveforms in Figure \ref{fig:ex3_Vabc_pscad} closely align with those from the full EMT simulation, indicating that the reconstructed waveforms from the TS-side voltage phasor in Figure \ref{fig:ex3_Vabc_psse} are inaccurate. The voltage difference curve calculated using Eq. (\ref{eq:eidx_integrand}) is presented in Figure \ref{fig:ex3_dvt}, where the non-zero values during fault significantly inflate the proposed error index, resulting in an unreasonably large $e_{\text{idx}} = 0.1806$ pu$\cdot$s. In contrast, applying the modified formula in Eq. (\ref{eq:eidx_int_mod}) results in a smaller value of $e^{\prime}_{\text{idx}} = 0.0925$ pu$\cdot$s, but it is not sufficiently small to be considered negligible. The significant difference between $e_{\text{idx}}$ and $e^{\prime}_{\text{idx}}$ indicates that the low positive-sequence voltage magnitude at the boundary bus strongly interferes with the proposed error index, and the fact that results from the TS side are inaccurate. Fortunately, the low boundary voltage magnitude---especially during short-circuit conditions---physically limits the electrical impact of the inaccurate TS region on the EMT region, as any erroneous dynamics from the TS side struggle to propagate through this low-voltage boundary bus. Consequently, the EMT-side voltage response remains accurate, as seen in Figure \ref{fig:ex3_Vm}. In this case, the large discrepancy between $e_{\text{idx}}$ and $e^{\prime}_{\text{idx}}$ does not necessarily indicate an inaccurate hybrid simulation; rather, it reflects an inaccurate representation of TS-side dynamics in the EMT model. It is important to note that this interface issue arises when the boundary bus experiences low positive-sequence voltage magnitudes, whether due to faults on the boundary bus or nearby buses. 


When $e_{\text{idx}}$ = $e^{\prime}_{\text{idx}}$, if they are both large, e.g., greater than $0.05$ pu$\cdot$s, it implies that the hybrid simulation contains significant error caused by the interface. If they are both small, then it implies that no significant errors are introduced by the hybrid interface.

\begin{figure}[tb!]
	\centerline{\includegraphics[width=0.9\columnwidth]{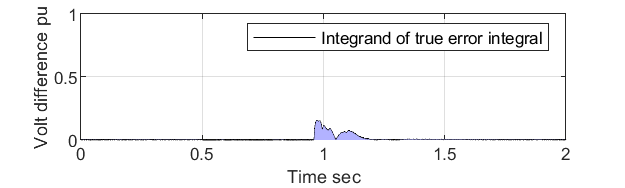}}
	\caption{True error of hybrid simulation in example 3}\label{fig:ex3_trueer}
\end{figure}

When $e_{\text{idx}} \neq e^{\prime}_{\text{idx}}$, we always have $e_{\text{idx}} > e^{\prime}_{\text{idx}}$ due to the inaccurate TS-side dynamics in the hybrid simulation falsely inflating $e_{\text{idx}}$. In this case, $e^{\prime}_{\text{idx}}$ is more reasonable, as depicted in Figures \ref{fig:erridx_lowV} and \ref{fig:ex3_trueer}. Therefore, $e^{\prime}_{\text{idx}}$ is used for evaluating the error by hybrid interface.


\subsection{Additional Tests on Faults}
\label{sec:faulttest}

\begin{figure}[tb!]
	\centerline{\includegraphics[width=0.8\columnwidth]{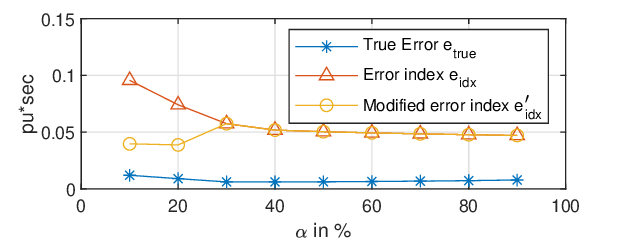}}
	\caption{Error vs. location of hybrid boundary for a balanced three-phase fault}\label{fig:add_alpha}
\end{figure}

\begin{figure}[tb!]
	\centerline{\includegraphics[width=0.8\columnwidth]{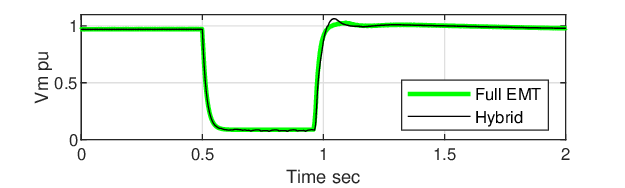}}
	\caption{Positive-sequence voltage magnitude of boundary bus with $\alpha=10\%$ in a balanced fault}\label{fig:add_alpha_Vm}
\end{figure}

\textbf{Balanced faults:} The first test in this subsection investigates the impact of the hybrid boundary location on balanced faults. The parameter $\alpha$ in Table \ref{tab:4busdata} is varied from $10\%$ to $90\%$ in $10\%$ increments, effectively changing the hybrid boundary's location. It is important to note that the total impedance between Bus 2 and Bus 4, as well as the short-circuit ratio at Bus 2, remain constant. A three-phase fault is introduced at Bus 2 at $t=0.5$ s and cleared at $t=0.96$ s. For each value of $\alpha$, the full EMT case and the hybrid case are constructed and simulated to calculate the true error, while the proposed error index is derived solely from the hybrid simulation results. The relationship between the error and the boundary location $\alpha$ is illustrated in Figure \ref{fig:add_alpha}. The results indicate that (i) the true error of the hybrid simulation, represented by the blue curve, remains consistently low---below $0.012$ pu$\cdot$s throughout the test, and (ii) the error index increases when $\|e_{\text{idx}}-e^{\prime}_{\text{idx}}\|$ is large, as seen at $\alpha=10\%$ in Figure \ref{fig:add_alpha}, which is primarily attributed to the low voltage at the boundary bus. A large error index $e_{\text{idx}}$ in this context does not imply an inaccurate hybrid simulation, as evidenced by the accurate voltage curves displayed in Figure \ref{fig:add_alpha_Vm}. Instead, $e^{\prime}_{\text{idx}}$ is better in this case.

Varying other factors, such as the short-circuit ratio at Bus 2 by adjusting $Z$ and the X/R ratio of the transmission network, yields results that further support the observations mentioned above. Due to space limitations, detailed results are omitted. Thus, we conclude that \emph{hybrid simulation consistently provides accurate results for balanced fault scenarios. A large error index indicates inaccuracies in the simulated dynamics from the TS side, while the EMT-side dynamics remain accurate.}

\begin{figure}[tb!]
	\centerline{\includegraphics[width=0.8\columnwidth]{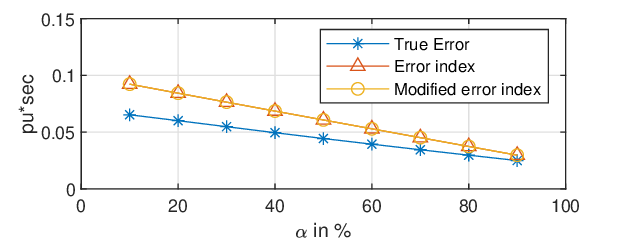}}
	\caption{Error vs. location of hybrid boundary for an unbalanced fault}\label{fig:add_alpha_1phfault}
\end{figure}

\begin{figure}[tb!]
	\centerline{\includegraphics[width=0.8\columnwidth]{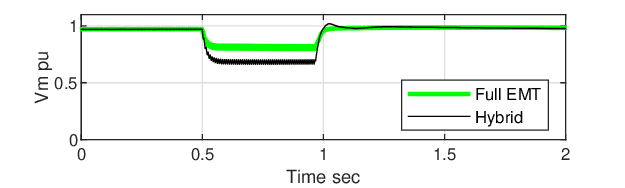}}
	\caption{Positive-sequence voltage magnitude of boundary bus with $\alpha=10\%$ in an unbalanced fault}\label{fig:add_alpha_Vm_1phfault}
\end{figure}

\begin{figure}[tb!]
	\centerline{\includegraphics[width=0.8\columnwidth]{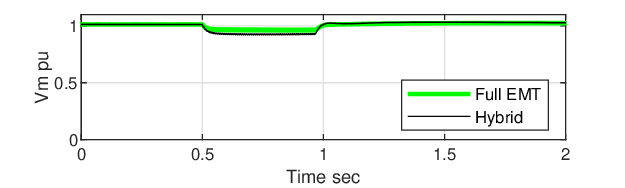}}
	\caption{Positive-sequence voltage magnitude of boundary bus with $\alpha=90\%$ in an unbalanced fault}\label{fig:add_alpha_Vm_1phfault_09}
\end{figure}

\textbf{Unbalanced faults:} The tests in this section are similar to the previous ones but focus on an unbalanced fault scenario. A bolted single-phase fault is introduced at Bus 2 at $t=0.5$ s and cleared at $t=0.96$ s. For each value of $\alpha$ shown in Figure \ref{fig:add_alpha_1phfault}, both full EMT and hybrid cases are constructed and simulated. The true error and the proposed error index are calculated and presented in Figure \ref{fig:add_alpha_1phfault}. The following observations are made: 

\begin{itemize}
\setlength\itemsep{0em}
\item The true error of the hybrid simulation is substantial when the EMT region containing the unbalanced fault is small; for instance, it reaches $0.065$ pu$\cdot$s when $\alpha=10\%$. This is reasonable, as a smaller EMT region exacerbates the unbalanced condition at the boundary bus, leading to a larger error in the hybrid simulation.
\item The difference $\|e_{\text{idx}}-e^{\prime}_{\text{idx}}\|$ remains consistently small, indicating that the low boundary bus voltage issue associated with the hybrid interface is not present in this case. With unbalanced faults, at least one or two phases typically remain healthy, ensuing that the positive-sequence voltage magnitude at the boundary bus stays above $0.5$ pu. Consequently, any observed large error index signifies an inaccurate hybrid simulation, such as $e_{\text{idx}}=e^{\prime}_{\text{idx}}=0.092$ pu$\cdot$s for $\alpha=10\%$. This is further illustrated in the comparison of positive-sequence voltage in Figure \ref{fig:add_alpha_Vm_1phfault}.
\item As the EMT region expands (i.e., with larger $\alpha$ values), the accuracy of hybrid simulation improves. For example, the enhanced positive-sequence voltage for $\alpha=90\%$ is depicted in Figure \ref{fig:add_alpha_Vm_1phfault_09}. According to this observation, a natural solution for reducing hybrid simulation errors caused by unbalanced faults is to expand the EMT region.
\end{itemize}

\section{Case Studies on Forced Oscillations}
\label{sec:num_FOs}

The goal of this section is to evaluate the accuracy of the hybrid simulation for low- and high-frequency dynamics. To achieve this, forced oscillations (FOs) are injected into the system due to the convenience in designing oscillatory cases with specific desired frequencies and amplitudes. This section first introduces the modeling of FOs and then presents tests on modulated FOs (MFOs) and superimposed FOs (SFOs). 

\subsection{Modeling of FOs in PSCAD}
\begin{figure}[tb!]
	\centerline{\includegraphics[width=0.75\columnwidth]{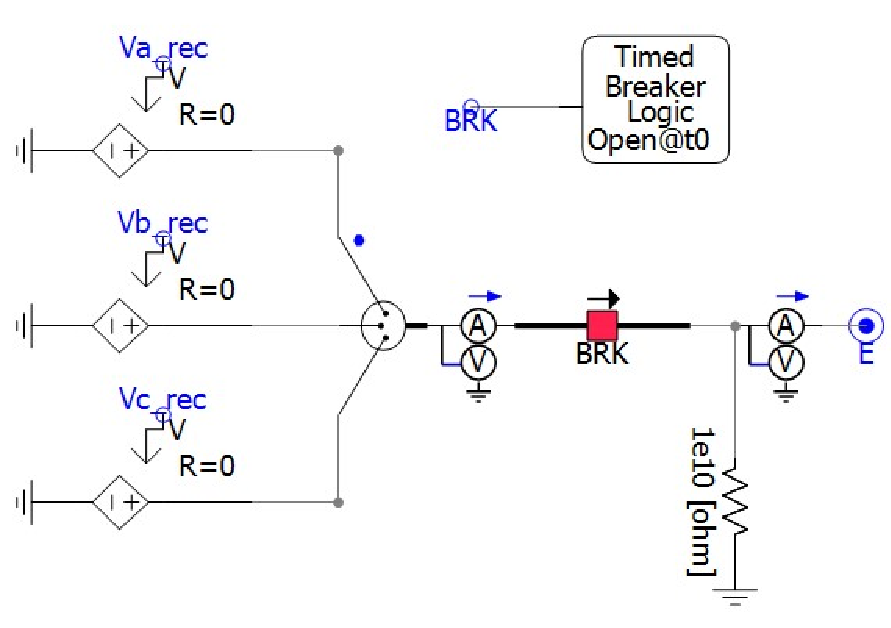}}
	\caption{A user-defined PSCAD component for injecting FOs}\label{fig:pscadfo}
\end{figure}

A user-defined component is created in PSCAD to model FOs. The electrical system, as shown in Figure \ref{fig:pscadfo}, consists of three single-phase, controlled ideal DC voltage sources, enabling the realization of any desired instantaneous voltage waveforms, along with a three-phase circuit breaker to facilitate the simulation initialization process. The control system for this component comprises two parts. The first part includes a phase-locked-loop that tracks the magnitude, phase, and frequency of the selected bus to which the FO component is connected. Under steady-state conditions, the tracked parameters are used to synthesize control signals that mimic the measured waveform, such as phase-a voltage as defined in (\ref{eq:sinwave}). This ensures that after the circuit breaker is closed, there is zero power flow between the connecting bus and the FO component. Once the FO component is activated, the second part of the control system introduces FOs in one of two ways, as defined in (\ref{eq:mfo}) and (\ref{eq:sfo}), where $V_{\text{fo}}$ and $\omega_{\text{fo}}$ are input parameters that can be specified to set the FO amplitude and frequency. 

\begin{figure}[tb!]
	\centerline{\includegraphics[width=0.75\columnwidth]{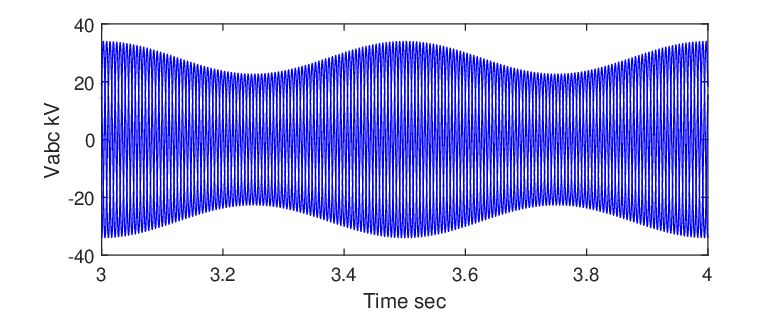}}
	\caption{A $2$-Hz MFO with $V_{\text{fo}}/V_{\text{m}}=0.2$}\label{fig:mfo}
\end{figure}

\begin{figure}[tb!]
	\centerline{\includegraphics[width=0.75\columnwidth]{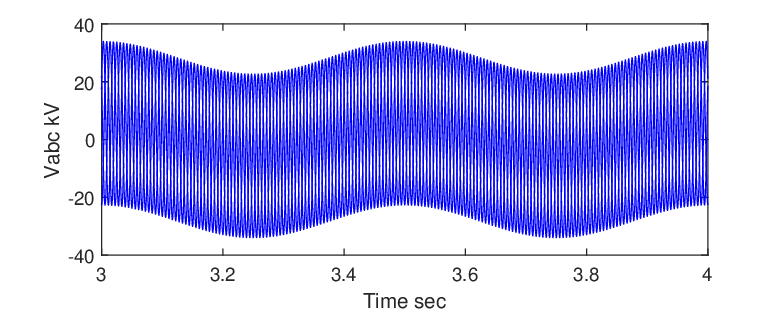}}
	\caption{A $2$-Hz SFO with $V_{\text{fo}}/V_{\text{m}}=0.2$}\label{fig:sfo}
\end{figure}

It is important to note that the FOs defined in (\ref{eq:mfo}) modulate an oscillation to the fundamental frequency and are thus referred to as MFOs, while those in equation (\ref{eq:sfo}) represent SFOs. Examples of a $2$-Hz MFO and a $2$-Hz SFO are illustrated in Figures \ref{fig:mfo} and \ref{fig:sfo}, respectively. When analyzing the frequency spectrum of the $2$-Hz MFO waveform, three frequency components are observed: $58$ Hz, $60$ Hz, and $62$ Hz. In contrast, the $2$-Hz SFO waveform exhibits only two frequency components: $2$ Hz and $60$ Hz. In fact, any MFO can be viewed as two superimposed frequency components that are symmetric about $60$ Hz and equal in amplitude.

\begin{equation}
    V_{\text{a,ss}}(t) = V_{\text{m}}\cos(\omega_{\text{syn}}t+\phi_{\text{a}}). \label{eq:sinwave}
\end{equation}

\begin{equation}
    V_{\text{a,mfo}}(t) = \big(V_{\text{m}} + V_{\text{fo}}\cos(\omega_{\text{fo}}t)\big)\cos(\omega_{\text{syn}}t+\phi_{\text{a}}) \label{eq:mfo}
\end{equation}

\begin{equation}
    V_{\text{a,sfo}}(t) = V_{\text{m}} \cos(\omega_{\text{syn}}t+\phi_{\text{a}}) + V_{\text{fo}}\cos(\omega_{\text{fo}}t) \label{eq:sfo}
\end{equation}

\subsection{Test of MFOs}
\begin{figure}[tb!]
	\centerline{\includegraphics[width=0.9\columnwidth]{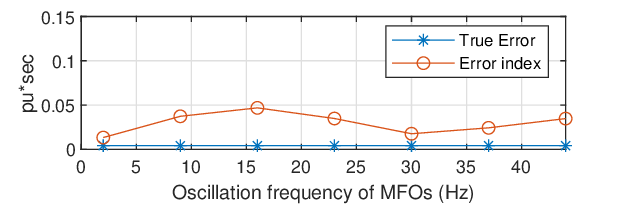}}
	\caption{Error vs. frequency in case of MFOs} \label{fig:err_mfos}
\end{figure}

\begin{figure}[tb!]
	\centerline{\includegraphics[width=0.9\columnwidth]{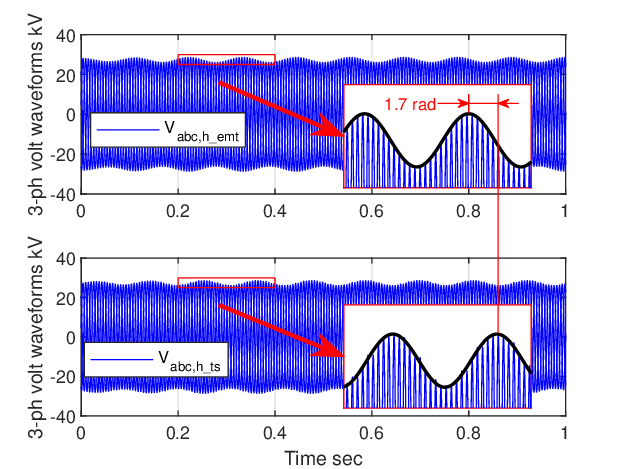}}
	\caption{Three-phase waveforms with $9$-Hz MFO: measured from EMT side vs. constructed from TS side} \label{fig:mfo_vabc_9}
\end{figure}

In the four-bus test system, the PSCAD FO component developed in the previous subsection is integrated into Bus 2 to introduce MFOs at frequencies of $2$ Hz, $9$ Hz, $16$ Hz, $23$ Hz, $30$ Hz, $37$ Hz and $44$ Hz, respectively. For each tested frequency, both full EMT and hybrid cases are constructed and simulated. The true error and the proposed error index are calculated and presented in Figure \ref{fig:err_mfos}. The blue curve consistently remains below $0.005$ pu$\cdot$s, indicating that the EMT-side results from the hybrid simulation are accurate for both low- and high-frequency dynamics. Figure \ref{fig:mfo_vabc_9} illustrates the voltage waveforms for the $9$-Hz MFO and demonstrates two key observations: (i) the $9$-Hz MFO can propagate from the EMT side into the TS side, and (ii) the oscillation phase of the $9$-Hz oscillation is mis-aligned by $1.7$ rad (or $97.4$ deg) across the hybrid interface. As shown by the orange curve in Figure \ref{fig:err_mfos}, the proposed error index can reflect the such phase mis-alignments, while a large mis-alignment in oscillation phase may disrupt and mislead any measurement-based oscillation analysis when studying modulated oscillations using hybrid simulations.


\subsection{Test of SFOs}
\begin{figure}[tb!]
	\centerline{\includegraphics[width=0.9\columnwidth]{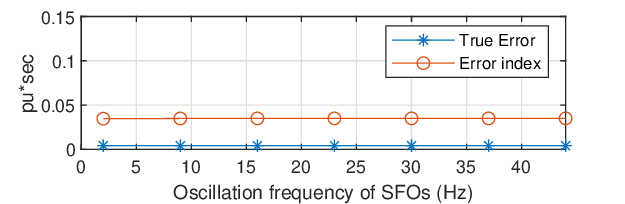}}
	\caption{Error vs. frequency in case of SFOs} \label{fig:err_sfos}
\end{figure}

\begin{figure}[tb!]
	\centerline{\includegraphics[width=0.9\columnwidth]{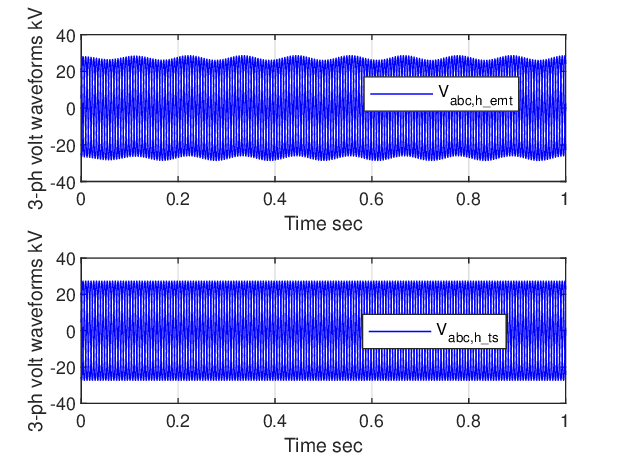}}
	\caption{Three-phase waveforms with $9$-Hz SFO: measured from EMT side vs. constructed from TS side} \label{fig:sfo_vabc_9}
\end{figure}

The similar test is repeated using SFOs, and the results are presented in Figure \ref{fig:err_sfos} and Figure \ref{fig:sfo_vabc_9}. It is observed that the $9$-Hz SFOs originating in the EMT region are blocked by the hybrid interface, rendering them completely invisible in the TS region. Consequently, the hybrid PSCAD-PSSE simulation may not be suitable for studying SFOs if any dynamic components from the TS side has significant participation. 


To conclude, EMT-TS hybrid interface can significantly distort oscillatory properties at the interface, causing errors in the TS-side dynamics at the oscillation frequency. For any dynamic components in the TS region that participate in the oscillations studied by the hybrid simulation, it is recommended to expand the EMT region to include these dynamic components. However, the proposed error index is not intended to determine whether a component significantly participates in a given oscillation mode. This is because oscillation properties may already be substantially distorted at the hybrid interface \cite{Ou2026PMU}, in which case a small error index computed from the distorted response does not reliably reflect the true system dynamics and should not be used to draw conclusions about modal participation.

\section{Three-Sequence Hybrid Interface}
\label{sec:3seq}

Previous tests have revealed that the EMT-TS hybrid simulation based on communicating only the positive sequence of the fundamental frequency component across the interface is largely accurate for balanced system conditions yet fall short of unbalanced conditions, e.g., unbalanced faults close to the boundary. Therefore, we propose a hybrid interface that can communicate 3-sequence information. Since commercial phasor simulation software like PSSE can run 3-sequence models, providing the sequence data are made available, the proposed solution is not only theoretically sound but is also practical.

\subsection{Modeling of Three-Sequence Hybrid Interface in PSCAD}

Since no three-sequence (3-seq) hybrid interface currently exists and developing one is technically challenging, an emulated 3-seq hybrid simulation case was created entirely in PSCAD for demonstration purposes. The four-bus test system is divided into two subsystems, connected at Bus 3 via the 3-seq interface, as shown in Fig. \ref{fig:FourBusDiagram}. Subsystem 1, comprising Buses 1, 2, and 3, is modeled identically to the EMT region in the original setup. Subsystem 2, which includes Buses 3 and 4, emulates TS dynamics and is also simulated in PSCAD. Three-phase voltage waveforms at Bus 3 in Subsystem 2 are converted into phasors before being sent to Subsystem 1. For practical applications, the 3-seq hybrid interface will need to be realized between EMT and TS platforms, with Subsystem 2 modeled in the TS domain, potentially containing thousands of buses. When Subsystem 1 receives the 3-seq voltage phasors from Subsystem 2, a two-step procedure is followed to reconstruct the three-phase instantaneous voltages: (1) the 3-seq voltage phasors are first transformed into three-phase voltage phasors using the inverse symmetrical components transformation \cite{wiki:symmetrical_components}, and (2) the three-phase voltage phasors are then converted, phase by phase, into three-phase instantaneous voltages \cite{wiki:phasor}.

As a first attempt of its kind, several technical difficulties were discovered that made the simulation numerically unstable. The following summarizes key issues and solutions for developing a numerically stable 3-seq hybrid interface.

\begin{itemize}
\setlength\itemsep{0em}
\item \textbf{Signals from EMT to TS side}. In the ETran Plus interface model, positive-sequence active and reactive powers are communicated from EMT side to the TS side. However, directly extending this solution to a three-sequence framework does not produce a functional hybrid simulation upon the occurrence of an unbalanced fault: (i) the negative- and zero-sequence powers change from zeros to non-zeros, while negative- and zero-sequence voltages remain zeros, resulting in a division-by-zero issue when computing current injection, (ii) unbalanced conditions from the EMT side cannot be accurately reflected in the TS side. Switching to sending current signals from EMT to TS side resolves this issue effectively.
\item \textbf{TS-side zero-sequence impedance}. The controlled current source on the TS side of the hybrid boundary injects currents received from the EMT side into the TS network. The stability of the zero-sequence response depends on the effective zero-sequence Thevenin impedance as seen by looking into the TS-side network from the boundary bus. If this impedance is excessively large, even a small zero-sequence current can produce a large zero-sequence voltage at the boundary bus and potentially lead to numerical instability. To mitigate this issue, it is sufficient to ensure that a finite zero-sequence grounding path exists from the boundary bus to the TS-side network, either directly or indirectly. This does not require grounding the entire TS zone, but instead simply avoids a floating zero-sequence network at the boundary. In realistic power system networks, such a condition is naturally satisfied because generators are effectively grounded either at the generator neutral or through the grounding of step-up transformers. In simulation studies, however, this grounding may not be properly represented EMT network models converted from TS models, where zero-sequence data are often absent or incorrectly assumed. As a result, unrealistically large zero-sequence impedances can be encountered at the hybrid boundaries.
\item \textbf{Phasor calculation}. Most phasor calculation methods process measurements from a specified time window to yield phasor magnitude and phase, with the phase typically corresponding to the first point of the time window. However, in the hybrid simulation, the current time should correspond to the last point of the window, i.e., the current time of the progressive simulation. Therefore, the initial phase should be adjusted to reflect the ending phase before transmitted to the other side of the hybrid interface.
\end{itemize}

After the abovementioned, a stable 3-sequence emulation environment is set up in PSCAD.

\subsection{Test of Unbalanced Fault}
\begin{figure}[tb!]
	\centerline{\includegraphics[width=1.0\columnwidth]{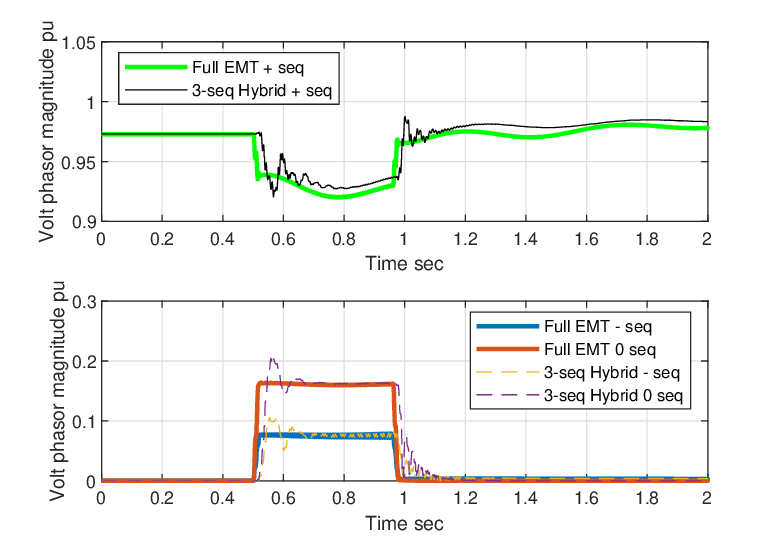}}
	\caption{Three-sequence voltage magnitude of boundary bus in a single-phase-to-ground fault event} \label{fig:3seq_1phfault}
\end{figure}

The same fault in Section \ref{sec:faulttest} was used to test the 3-seq interface. A single-phase fault with a $4$-$\Omega$ fault impedance is introduced at Bus 2 at $t=0.5$ s and cleared at $t=0.96$ s. Unlike Figure \ref{fig:add_alpha_Vm_1phfault}, where significant discrepancies arise due to the positive-sequence based hybrid interface, the proposed 3-seq hybrid interface additionally facilitates the exchange of negative- and zero-sequence quantities. This enhancement not only captures system's negative- and zero-sequence responses, but also improves the accuracy of the system's positive-sequence response. As illustrated in Figure \ref{fig:3seq_1phfault}, major dynamics are successfully captured by the 3-seq hybrid simulation. 

The fast spurious oscillations observed in Fig. \ref{fig:3seq_1phfault} are primarily caused by the one-time-step communication delay inherent in the three-sequence interface implemented in this study, where the signals between the EMT and TS domains are being exchanged. This delay introduces numerical artifacts that can inadvertently result in high-frequency oscillations. In comparison, hybrid interface models from commercial software such as ETRAN Plus use equivalent network models on each side of the interface, enabling partial, instantaneous interactions within the same time step and thereby improving numerical stability and accuracy. Developing strategies to mitigate interface errors, such as using the network equivalencing approach \cite{2016electranix:etranplus}, is an important future task.

Note that changing the hybrid interface from positive-sequence-only to three-sequence does not result in any noticeable increase in simulation runtime. This is because hybrid simulations are always bottlenecked by the EMT region, while the computational demands of the TS region and the interface are relatively insignificant.


\section{Conclusions}
\label{sec:con}
This paper evaluates the accuracy of the PSCAD-PSSE ETran Plus interface for hybrid electromagnetic transient-transient stability (EMT-TS) simulation and proposes an error index to capture the interface-induced inaccuracies. The calculation of the index is independent of full EMT simulation results, and relies solely on hybrid simulation results.

Extensive testing shows that hybrid EMT-TS simulations are generally accurate for balanced faults but may become inaccurate under unbalanced fault conditions. For balanced faults, the proposed error index can become falsely large when the boundary voltage is low. This is due to incorrect dynamics introduced by the interface and TS region, even though the EMT-side dynamics remain fairly accurate. To support a practical verification step, a revised error index is introduced to suppress the influence of false dynamics from TS side of the interface. Comparing it with the original index helps determine the presence of inaccuracies caused by the interface and TS region. In the case of unbalanced faults, the original and revised error indices are always equal, and high values reliably indicate simulation errors. In practice, engineers can begin with an initial hybrid boundary and expand the EMT region when the error index exceeds an acceptable threshold.

Tests on modulated and superimposed forced oscillations (FOs) reveal that hybrid interface distorts oscillation phases in modulated FOs and blocks superimposed oscillations at the boundary. Thus, hybrid simulation is unsuitable for analyzing oscillations where TS-side dynamics have large participation.

For unbalanced faults, expanding the EMT region to improve accuracy may not always be preferred. To address this, this paper introduces a three-sequence (3-seq) hybrid interface model that facilitates the exchange of negative- and zero-sequence quantities alongside positive-sequence quantities. Results demonstrate that the proposed 3-seq interface significantly enhances the accuracy of hybrid EMT-TS simulations for unbalanced faults.

Future work will include the 1) determination of threshold in the studies with different fault types, 2) application of the proposed error index for determining hybrid boundaries in studies involving realistic grid network models, and 3) selection of the stopping criteria for the expansion of the EMT region.

\section*{Acknowledgment}
The authors would like to acknowledge Garth Irwin, Chaminda Amarasinghe and Andrew Isaacs from Electranix for the helpful discussions.

\bibliographystyle{ieeetr}
\bibliography{references}

\end{document}